\documentstyle[11pt]{article}

\def\beq{\begin{equation}}
\def\eeq{\end{equation}}

\begin{document}
\begin{center}
{\large\bf Transverse momentum distributions and
their forward-backward correlations in
the percolating colour string approach}\vspace{0.4 cm}

M.A.Braun$^a$ and C.Pajares$^b$\vspace{0.2 cm}

{\small\it $^a$ Department of High
Energy Physics,
University of St. Petersburg,\\ 198904 St. Petersburg, Russia\\
  $^b$ Departamento
de F\'{\i}sica de Part\'{\i}culas, Universidade de Santiago de
Compostela,\\ 15706-Santiago de Compostela, Spain}
\end{center}

{\small The forward-backward correlations in the $p_T$ distributions,
which present a clear signature of  non-linear effects in particle production,
are studied in the model of percolating colour strings. Quantitative
predictions are given for these correlations at SPS, RHIC and LHC energies.
Interaction of strings also naturally explains the flattening of $p_T$ 
distributions and increase of $\langle p_T\rangle$ with energy and atomic number
for nuclear collisions}.\vspace{1cm} 

The study of $p_T$ distributions in hadronic and nuclear reactions at
high energies offers a unique opportunity to observe  non-linear effects
in high-density
nuclear matter. Indeed in a simple picture in which particle production
goes via formation of several independent emitters, colour strings
stretched between the projectile and target, the $p_T$ distribution
is independent of the number of strings and coincides with the
$p_T$ distribution from a single colour string. As a result, in this picture
the observed $p_T$ distribution does not depend 
on the energy nor on the atomic number of the colliding particles.
Experimentally this prediction is fulfilled only in a very crude manner.
In fact the average $\langle p_T\rangle$  grows with energy and also with
the atomic number of the participants (the "Cronin effect"). This behaviour
clearly shows that particles are produced by different strings not
independently. In other words, strings interact with each other and
these non-linear effects result in the experimentally observed behavior
of the $p_T$ distributions.

Some time ago the authors introduced a simple model which introduces
the interaction between strings via their fusion and percolation [1,2].
In this note we study the $p_T$ distributions in this model.
We show that the model describes both qualitatively and quantatively
the behaviour of the $p_T$ distributions of the produced particles
with the growth of both the energy and atomic number of the participants.
We also study the forward-backward correlations (FBC) in the $p_T$
distributions. We find that their mere existence  is a signature of
non-linear effects in particle production. Our model predicts a very
concrete form of these correlations, which can be tested in present-day
and future experiments.

We start from the particle spectra produced by a single colour string.
The standard idea exploited in their description is that the observed particles
are formed via quark-antiquark pair emissions by the colour field
of the string. According to [3], the probability rate for this production then
has a Gaussian form as a function of $p_T$.
As soon as a $q\bar q$ pair is created, the colour charge $Q$ of the string
diminishes, so that at the next step the production rate is different
(although also of the Gaussian form in $p_T$). Detailed
caluclations show that the resulting $\langle p_T^2\rangle$ of particles
produced after the complete break-up of the string is proportional to $Q$
[4]
\beq
\langle p_T^2\rangle_{Q} =Q\langle p_T^2\rangle_1,
\eeq
where $\langle p_T^2\rangle_1$  is the average $p_T^2$ of particles
produced by a "single" string, created by colour $Q_0$
corresponding to a single $q\bar q$ pair.
In future we measure  all colours in units $Q_0$,
which is equivalent to putting
$ Q_0=1$.
To simplify the notation we also omit the subindex $T$ in the momentum
distribution, since we shall only consider the transverse momenta.

The Gaussian distribution in $p$ can be however theoretically
supported only for infinitely high energies, which correspond to
a string infinitely long in rapidity. For realistic strings with
a finite length in rapidity one expects corrections due to energy
conservation.  
Also fluctuations of the string tension may change the form of the
$p$-distribution, as advocated in [5]. Finally an evident
 restriction comes at large $p$ where the
hard collision mechanism is expected to be responsible for particle
creation. As a result one expects the $p$ distribution to have a 
power behaved tail.

A realistic form of the $p$ distribution corresponding to a 
single colour string can be extracted
from the experimentally measured one in 
$p\bar p$ soft collisions at 630 and 1800 GeV/c [6]. Assuming that the effects
of string interactions are small for $p\bar p$ interactions, we may take that
the measured distribution coincides with the one for 
a single  ordinary ($Q=1$) string. This
distribution has a form 
\beq
w_1(p)=\frac{(k-1)(k-2)}{2\pi p_0^2}\frac{p_0^k}{(p+p_0)^k}.
\eeq
Comparing the distributions at 630 and 1800 GeV and also taking into account the
behaviour of the minimum bias distributions in all the energy interval from 63
to 1800 GeV [7]  we parametrize 
\beq
p_0=2 {\rm GeV/c},\ \  k=19.7-0.86 \ln E,
\eeq
where $E$ is the c.m. energy in GeV.
With (2)  the averages $\langle p\rangle$ and $\langle p^2\rangle$ are given  by
\beq
p_1\equiv\langle p\rangle_1=p_0\frac{2}{k-3},\ \ \ \langle
p^2\rangle_1=p^2_0
\frac{6}{(k-3)(k-4)}. 
\eeq

Passing to the string with colour $Q$, to satisfy (1), we change
$p^2_0\rightarrow Qp_0^2$, so that our  distribution
corresponding to the string with colour $Q$ is
\beq
w_Q(p)=\frac{(k-1)(k-2)}{2\pi Qp_0^2}\frac{(p_0\sqrt{Q})^k}{(p+p_0\sqrt{Q})^k}.
\eeq
We stress that this distribution only refers to the soft part of the spectrum,
which supposedly is generated by the string decay. The observed spectrum also
contains a contribution from hard events (with a  produced 
cluster having $p>$1.1 GeV/c, in the definition of [6]).

Our picture for the high-energy particle production consists in
assuming that in the collision several colour strings are created
which may overlap in the transverse space. In the overlap area the
colour fields of the strings add algebraically. Due to the
vector character of color charge, the resulting colour squared
of the overlap area is just a sum of the colour squared of the
overlapping strings. Thus in the overlap of $n$ strings a new
colour string is formed corresponding to colour
$ Q_n=\sqrt{n}$.
The fact that the colour of the overlapping strings is proportional
to the square root of their number and not to their number has
an immediate  consequence of damping the multiplicities of the
produced particles by a factor of the order three at the LHC energies [8].
Here we study its influence on the $p$-distribution.

To find the latter we have to know the distribution in the areas of overlaps of
$n$ strings, which gives the weights with which different overlaps
contribute to particle production. It has been shown in [8] that in the
"thermodynamic limit", that is for an idealized system with a very large
total interaction area $S$ and corresponsingly large number of strings
$N$, the properties of the system and the overlap
distribution in particular are governed by a dimensionless parameter
\beq
\eta=\sigma_0\frac{N}{S}=\sigma_0\rho.
\eeq
Here $\sigma_0$ is an area of a single string and $\rho$ the string density.
Note that at $\eta>\eta_c\simeq 1.12-1.20$ the percolation phase
transition occurs, most of the space being occupied by a single cluster
formed by many overlapping strings.

It has been shown in [8] that in the thermodynamic limit the distribution
in the overlaps areas with different $n$ follows the Poisson law
 with an average value equal to $\eta$:
\beq
\lambda(n)=\frac{S_n}{S}=a(\eta)\frac{\eta^n}{n!}
\eeq
with
 $a=\exp (-\eta)$.
Here $S_n$ is the total area in which exactly $n$ strings overlap.
Eq. (7) is valid for $n=0,1,2...$, $\lambda(0)$ giving the part of the area
in which there are no strings at all. Evidently the latter
part does not produce particles. So
we are only interested in the relative contribution to particle
production of overlaps with $n=1,2,...$. This will be given by (7)
with a different normalization factor $a=1/(\exp\eta-1)$.

The overall $p$ distribution $P(p)$ is obtained by 
 convoluting the distribution (5) in $p$ for 
fixed $Q$ and (7) in $n$ and taking into account that for an overlap of
$n$ strings $Q=\sqrt{n}$:
\beq
P(p)=\sum_{n=1}\lambda(n)w_{\sqrt{n}}(p).
\eeq

As a consequence of (1) we find  
\beq
\langle p\rangle=p_1a(\eta)\sum_{n=1}n^{1/4}\frac{\eta^n}{n!}.
\eeq
For small $\eta$ this gives
$
\langle p\rangle/p_1=1+0.094\eta.
$
The behaviour of $\langle p\rangle/p_1$ for  $0.5<\eta<4$  can also be well
described by a linear dependence
$1+0.098\eta$.

So even with a fixed average $p_1$  for a single string, the overall
average grows with $\eta$. On the other hand, $\eta$ grows both with the
energy $E$ and the atomic number $A$ of the  colliding particles
[8], so that fusion of strings by itself leads to the growth of the average
transverse momentum with $E$ and $A$. With the distribution (2)
the average $p_1$ also rises with energy 
from 0.28 GeV/c at 19.4 GeV to 0.43 GeV/c
at 5500 GeV. This rise has to be combined with that due to the growth of $\eta$
for collisions with nuclei. Using our earlier calculations [8] we find the 
values of $\eta$  for central p-PB and Pb-Pb collisions at c.m. energies 19.4,
200 and 5500 GeV shown in the second and fourth columns of the Table. In
the third and fifth columns we present the corresponding values of $\langle
p\rangle$ which follow from Eq (9) with the distribution (5). Comparing these
values with the experiment, one should remember that they refer only to
the soft part of the spectra.

The form of the $p$ distribution in p-Pb and Pb-Pb collisons
corresponding to Eq.(8) is
shown in Fig. 1. With the growth of $E$ and $A$
the high $p$ tail is strongly enhanced.
This corresponds to the well-known Cronin effect.
In Fig. 2 we present the
ratio of the distributions for the  p-Pb and Pb-Pb to p-p reactions.
They are well compatible with the experimental ones [9].

Now we pass to the FBC in the $p$-distributions.
They can be studied from the
observation of the average $p$ in the backward hemisphere
$\langle p_B\rangle_{p_F}$ for events with given $p=p_F$ in the forward
hemisphere.
In absence of any interaction between colour strings
(independent colour string model) the average $p$ in both hemispheres
evidently is identical with this average for a single string.
Then
\beq
F(p_F)\equiv\langle p_B\rangle_{p_F}/p_1=1.
\eeq
Of course, one should have in mind that correlations between different
strings are imposed not only but their fusion and percolation, but also
on purely kinematical grounds, due to energy conservation. However this
latter effect diminishes with energy and becomes neglegible in the
mid-rapidity region at large enough energy. Modulo this kinematical
effect, the mere existence of FBC, that is, the difference of the
right-hand side of Eq. (10) from unity is a clear signature of a dynamical
interaction  between strings. In our percolation string model
function $F$ can be found explicitly.

With two different observables $p_B$ and $p_F$ we introduce the
corresponding probability $w(p_F,p_F)$ which generalizes Eq. (5) 
and shows the probability to find the observed particle with transverse
momenta $p_F$ and $p_B$ in the forward and backward hemispheres.
Technically it can be found from the inclusive cross-section
$2Ed^3\sigma/d^3p$
integrated over the angles in the forward or backward hemispheres
respectively and properly normalized.
Our starting point will be an assumption that there is no correlation
between emission of particles in the forward and backward hemispheres
for a single string:
\beq
w_Q(p_F,p_B)=w_Q(p_F)w_Q(p_B),
\eeq
where each of the functions $w$ on the right-hand side is given by
Eq. (5).  Then for a single string of colour $Q$
\beq
\langle p_F\rangle_Q=\langle p_B\rangle_Q=\langle p\rangle_Q=\sqrt{Q}p_1.
\eeq
Passing to the system of percolating strings with different overlaps we
find the final distribution in $p_F$ and $p_B$ as a suitable
generalization of (8)
\beq
P(p_F,p_B)=\sum_{n=1}\lambda(n)w_{\sqrt{n}}(p_F,p_B).
\eeq
Integrating this  expression over $p_F$ one  obtains
the distribution in $p_F$:
\beq
P(p_F)=\int d^2p_BP(p_F,p_B)=
\sum_{n=1}\lambda(n)w_{\sqrt{n}}(p_F)=P(p).
\eeq
As expected, it coincides with the overall distribution (8).

The conditional probability to see a particle with momentum $p_B$ in 
the backward
hemisphere, provided one observes a particle with momentum $p_F$ in the
forward hemispere, is given by the ratio
\beq
P(p_B)_{p_F}=\frac{P(p_F,p_B)}{P(p_F)},
\eeq
so that the average value  of any observable $A(p_F,p_B)$ for a given fixed
$p_F$ is given by
\beq
\langle A\rangle_{p_F}=P^{-1}(p_F)\int d^2p_BA(p_F,p_B)P(p_F,p_B).
\eeq

Taking
$ A(p_F,p_B)=p_B$ we obtain
for the function $F$ in (10) 
\beq
F(p_F)=
\frac{\sum_{n=1}n^{1/4}\lambda(n)w_{\sqrt{n}}(p_F)}
{\sum_{n=1}\lambda(n)w_{\sqrt{n}}(p_F)}.
\eeq
Its
difference from unity measures the FBC in the transverse momentum distributions.

At small $\eta$ we find 
\beq
F(p_F)=1+0.0669\,\eta \frac{(1.189(p+p_0))^k}{(p+1.189p_0)^k}.
\eeq
So  positive correlations follow (as expected), which
grow with the momentum from a non-zero value to saturate at a certain larger
value depending on the energy (via $k$).

The behaviour of $F(p_F)-1$ for the reactions p-Pb and Pb-Pb at different
energies is illustrated in Fig.3. The FBC rise with energy and atomic number,
their magnitude well allowing for the experimental observation.

Summarizing, we have shown that interaction of strings predicts sizable FBC
in the transverse momenta, which can be tested in the forthcoming experiments.
Also the dependence of the $p$-spectra on $E$ and $A$ is naturally explained
for collisions with nuclei.

This work has been done under the contract AEN99-0589-C02-02 from CICYT of
Spain.
\vspace{1 cm}
 
References
\vspace{0.5 cm}

[1] M.A.Braun and C.Pajares, Phys. Lett. {\bf B287}  154 (1992);
Nucl. Phys. {\bf B390} 542,549  (1993);
N.S.Amelin, M.A.Braun and C.Pajares, Phys. Lett. {\bf B306} 312 (1993);
Z.Phys. {\bf C63}  507 (1994).

[2] N.Armesto, M.A.Braun, E.G.Ferreiro and C.Pajares,
Phys. Rev.Lett. {\bf 77} 3736  (1996).

[3] J.Schwinger, Phys. Rev.{\bf 82} 664  (1951);
 T.S.Biro, H.B.Nielsen and J.Knoll, Nucl. Phys. {\bf B245} 449  (1984).

[4] A.Bialas and W.Czyz, Nucl. Phys. {\bf B267} 242  (1986) .

[5] A.Bialas, hep-ph/9909417

[6] CDF Collab, F.Rimondi, Talk presented at IX Int. Workshop on Multiparticle
Production, Torino (2000), to appear in Nucl. Phys. {\bf B} (Proc).

[7] UA1 Colab, G. Arnison et al., Phys. Lett. {\bf 118 B} (1982) 167;
CDF Collab., F.Abe et al., Phys. Rev. Lett., {\bf 61} (1988) 1819.

[8] M.A.Braun and C.Pajares, to be published in Eur. Phys. J. C;
N.Armesto and C.Pajares, Int. J. Mod. Phys. {\bf A 15} (2000) 2019

[9] J.Schukraft, in Proc. Int. Workshop on Quark-Gluon Plasma
Signatures, Strasbourg, France (1990)
\vspace{1 cm}

Figure captions
\vspace{0.5 cm}

Fig. 1. Transverse momentum distributions in the
reactions p-Pb and Pb-Pb at different c.m.energies
(normalized to unity). Curves 1,2 and 3 correspond to p-Pb, and curves
4,5 and 6 to Pb-Pb at 19.4, 200 and 5500 GeV respectively.

Fig. 2. Ratio of $p_T$ -distributions for the reactions p+Pb and Pb-Pb to p-p
at c.m. energies 19.4, 200 and 5500 GeV. For p-Pb the curves go down with energy.
For Pb-Pb the slope rises with energy

Fig. 3. The FBC parameter F-1 (Eq. (23)) as a function of $p_F$
for the reactions p-Pb and Pb-Pb at c.m. energies 19.4, 200 and 5500 GeV. 
For p-P the curves rise with energy.
\vspace{1cm}

Table
\vspace{0.5 cm}
\begin{center}
\begin{tabular}{|l|r|r|r|r|}\hline
   &\multicolumn{2}{c|}{p-Pb}&\multicolumn{2}{c|}{Pb-Pb}\\\hline
Energy (GeV)& $\eta$& $\langle p_T\rangle$ (GeV/c)&$\eta$&$\langle
p_T\rangle$ (GeV/c)\\\hline
19.4& 0.53& 0.30   &1.19& 0.32 \\\hline
200 & 0.60& 0.35   &1.82& 0.39 \\\hline
5500& 0.76& 0.46   &3.54& 0.58 \\\hline
\end{tabular}
\end{center}
\vspace{1 cm}

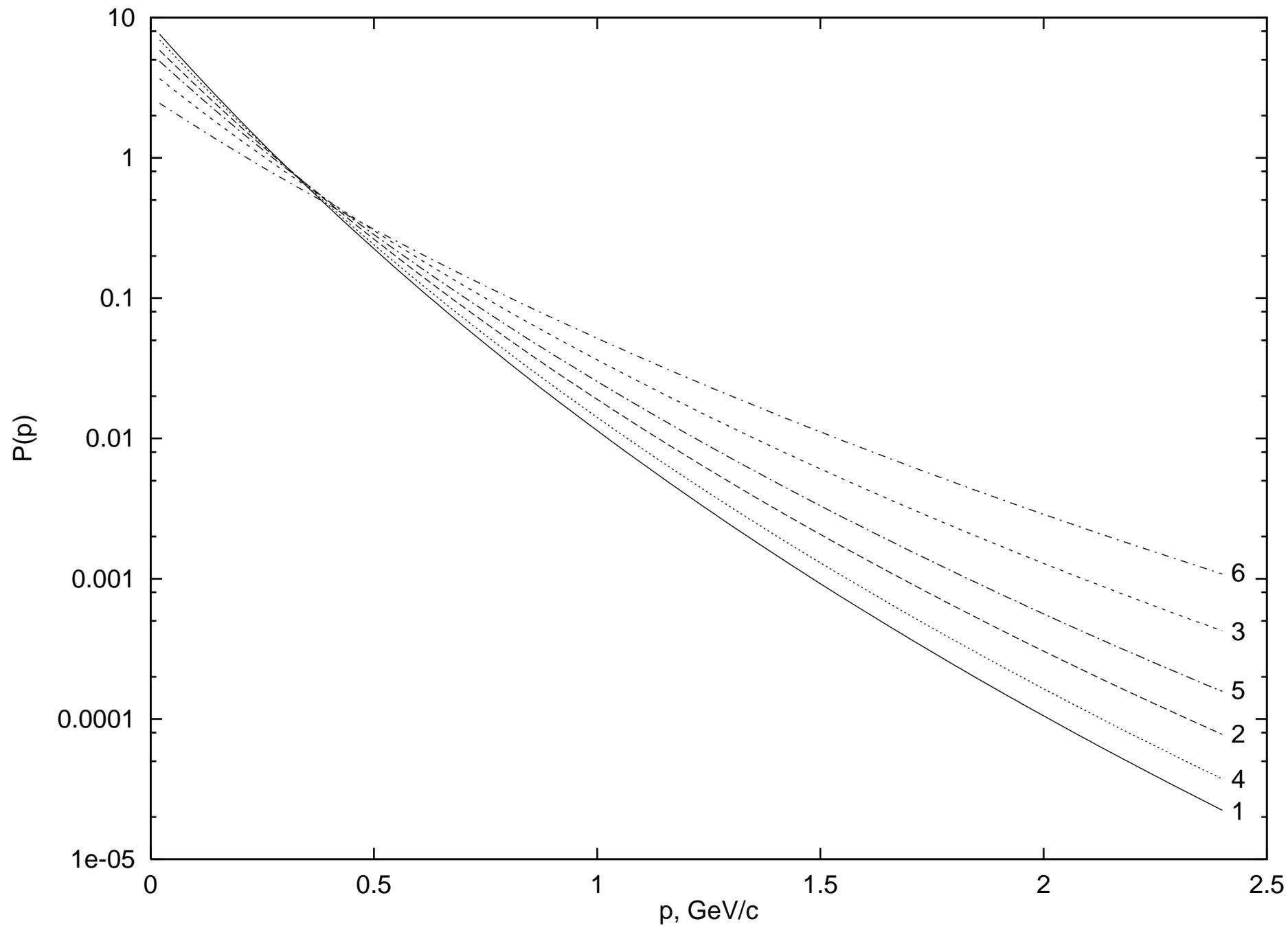
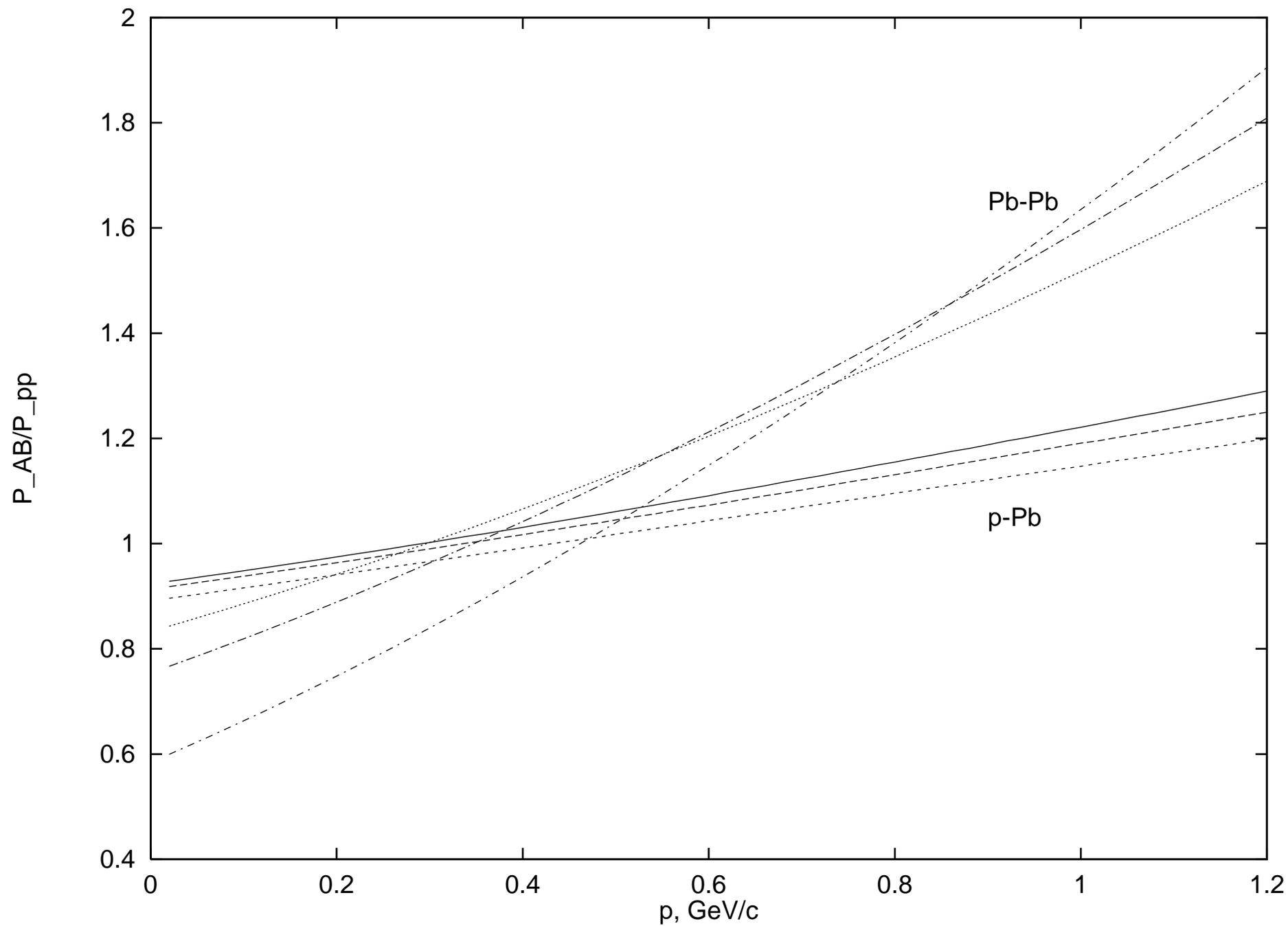
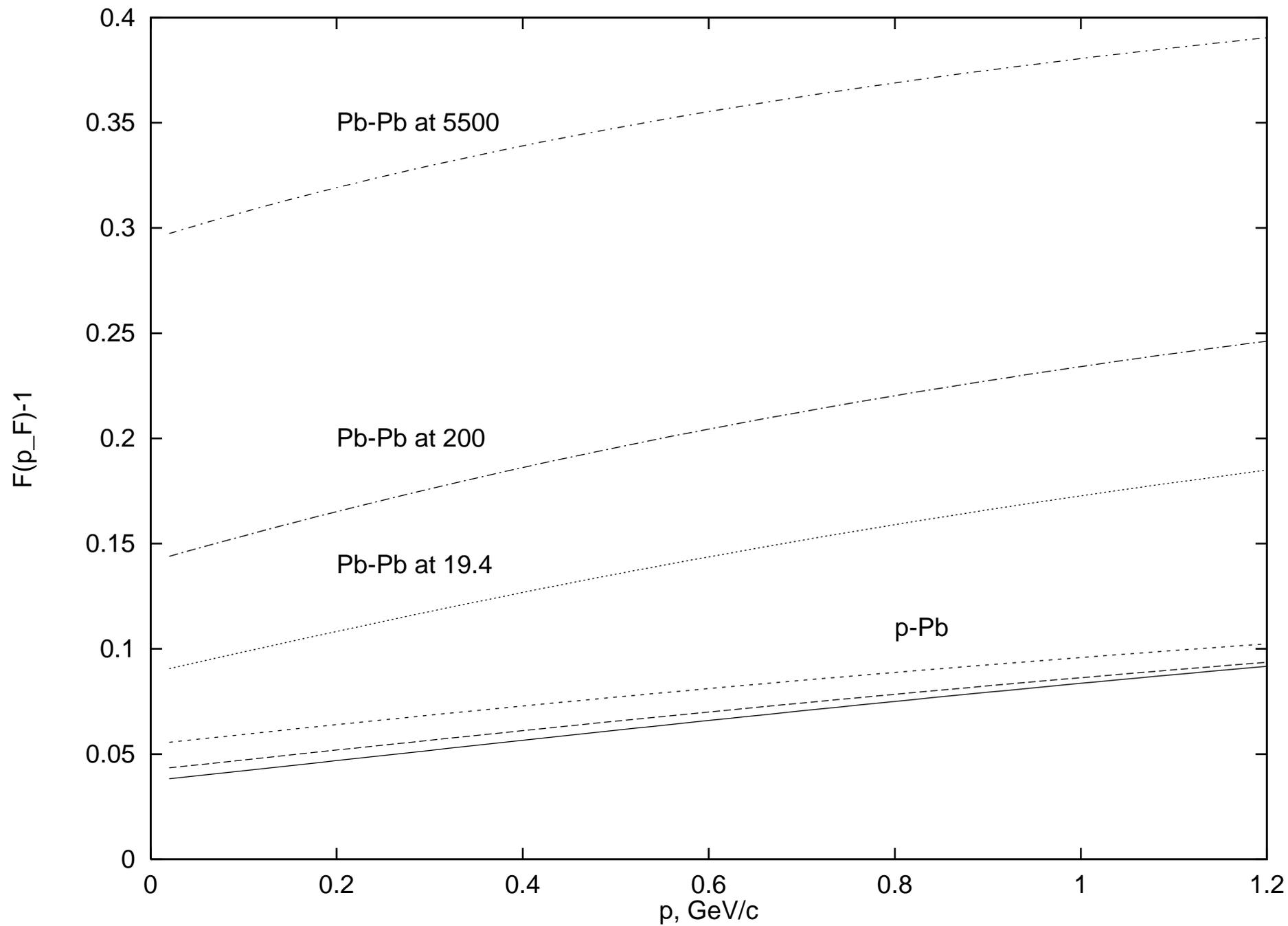
\end{document}